\documentclass[pra,amssymb,twocolumn,tightenlines,showpacs]{revtex4-1}

\usepackage{graphicx}
\usepackage{amsmath}
\usepackage{verbatim}
\usepackage{subfigure}
\usepackage{hyperref}
\usepackage{epstopdf}

\begin{document}

\title{Demonstration of quantum error correction for enhanced sensitivity of photonic measurements}

\author{$^*$ L. Cohen}
\thanks{L.C. and Y.P. made an equal contribution to this work}
\affiliation{Racah Institute of Physics, Hebrew University of
Jerusalem, Jerusalem 91904, Israel}
\author{Y. Pilnyak}
\affiliation{Racah Institute of Physics, Hebrew University of
Jerusalem, Jerusalem 91904, Israel}
\author{D. Istrati}
\affiliation{Racah Institute of Physics, Hebrew University of
Jerusalem, Jerusalem 91904, Israel}
\author{A. Retzker}
\affiliation{Racah Institute of Physics, Hebrew University of
Jerusalem, Jerusalem 91904, Israel}
\author{H. S. Eisenberg}
\affiliation{Racah Institute of Physics, Hebrew University of
Jerusalem, Jerusalem 91904, Israel}


%
%
%

\begin{abstract}
The sensitivity of classical and quantum sensing is impaired in a
noisy environment. Thus, one of the main challenges facing sensing
protocols is to reduce the noise while preserving the signal.
State of the art quantum sensing protocols that rely on dynamical
decoupling achieve this goal under the restriction of long noise
correlation times. We implement a proof-of-principle experiment of
a protocol to recover sensitivity by using an error correction for
photonic systems that does not have this restriction. The protocol
uses a protected entangled qubit to correct a single error. Our
results show a recovery of about $87\%$ of the sensitivity,
independent of the noise rate.
\end{abstract}

\maketitle

\section{Introduction}
Error correction is an essential ingredient in classical
computation and communication and is currently used in many state
of the art technologies. Shor's discovery of an algorithm
\cite{Shor95} to break the RSA cryptosystem showed that a quantum
computer is a promising system for tackling hard computational
problems. However, it was not clear if even conceptually a quantum
computer could be constructed. The reason for the large doubt, is
that even a small error in every computational step would
accumulate quite swiftly to a large error. It implies that the
future quantum computer could solve only small and probably
trivial problems. Remarkably, the theory of quantum fault
tolerance \cite{Knill, Aharonov, Kitaev, Aliferis} showed that
this intuition is wrong. Actually, what is needed for the
calculation of a quantum computer to give the right result with a
small probability of failure is that every gate operation will
fail with a small probability below a certain threshold. Put
simply, a gate error below the threshold is effectively as good as
no error at all.  To date, many quantum error correction schemes
have been suggested
\cite{Gottesman09,Shor95,Steane,Laflamme,Gottesman,Aharonov,Kitaev}
and implemented \cite{Schindler11,Lassen13,Waldherr,Nigg14,Kelly}.
The predicted ability of a fault tolerant quantum computer to
reach any desired precision raises a question whether this
remarkable precision could be used for precise measurements.

Quantum metrology \cite{Giovannetti11}; i.e., enhanced metrology
using quantum mechanics, is a well developed field with
applications to photonic interferometry \cite{Aasi13,Israel13},
magnetometry \cite{maze08,Balas08,Budker_Review} and atomic
spectroscopy \cite{Schmidt05}. However, the quantum advantage
rapidly degrades as noise takes its toll \cite{huelga}. Therefore,
to maintain this advantage in a noisy environment, error
correction needs to be applied. Several theory papers have been
devoted to this issue that deal with a variety of noise types
\cite{Alex14, Lu15, Yuan15, Demkovich14, Kessler14,Dur,Ozeri} and
there has been one implementation with NV centers in diamond
\cite{thomas} using a carbon nuclear spin as the redundant qubit
and a related implementation \cite{paola} using the NV nuclear
spin as the redundant qubit.

In this work, we implement a combination of error correction and
sensing for the first time using a photonic platform. The protocol
we implement uses only linear optical elements. The error
correction relies on the use of a protected entangled photon qubit
and a gate which is realized by post-selection. Not only the phase
information, which we are interested in, is preserved, but the
entire quantum state as well. Due to the use of post-selection,
each error correction cycle is only $50\%$ viable; however, this
setup provides a proof of principle experiment that could have
been achieved with deterministic gates which are currently being
developed \cite{kimble95,lukin15,arno14,barak14}.

\section{Theoretical background}

Let us consider a quantum state evolving under a Hamiltonian which
depends on an unknown parameter, which we want to measure. The
state changes continuously as a function of the parameter as it
propagates through a noisy path vulnerable to a single type of
error, either a bit-flip or a phase-flip \cite{Alex14,Kessler14}.
If the data are integrated without an error correction,
sensitivity drops as a result of cancellation between measurements
with and without the error. However, if an error correction
protocol is applied much faster than the time at which the
deleterious effects of the noise become significant, the
sensitivity level can be maintained. This scenario can be
simulated by repeating a series of parameter probing and noise
operations many times. Instead of measuring once over a time $T$,
as in the usual sensing scheme, a measurement of duration
$\frac{T}{N}$ is repeated $N$ times, where between each
measurement the error is corrected.


We present and implement the protocol by the optical realization
of qubits using the polarization degree of freedom of photons, and
denote the vertical (horizontal) linear polarization as
$|V\rangle$ $(|H\rangle)$. The birefringence phase, denoted as
$\theta$, is the unknown parameter. The scheme is composed of a
regular interferometric setup for measuring a birefringent phase
\cite{Cohen14}. However, in the presented model, the measurement
of the birefringent phase is also associated with a noisy
environment that flips the qubit. Although we chose to correct
bit-flips, phase-flip errors could have been corrected instead
with a minor addition of polarization rotators.

\section{Experimental setup and state evolution}

\begin{figure}[tb]
\centering\includegraphics[angle=270,width=86mm]{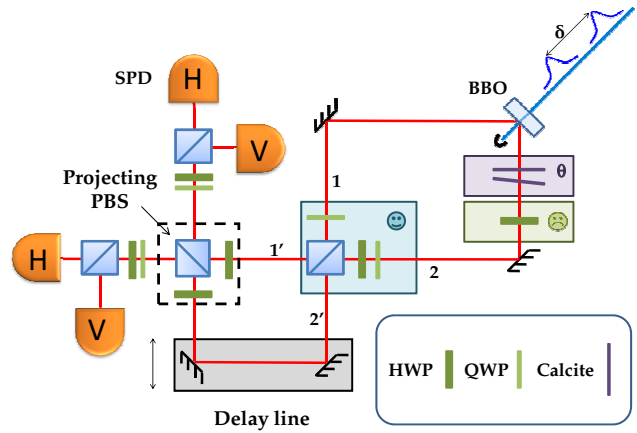}
\caption{\label{ExperimentalSetup}(color online). The experimental
setup. $780\,nm$ entangled photons are generated by a
non-collinear type-II parametric down-conversion from $2\,mm$
thick $\beta$-BaB$_2$O$_4$ (BBO) crystal using a $390\,nm$ doubled
Ti:Sapphire pulsed laser \cite{Kwiat95}. One photon measures a
birefringence phase ($\theta$ in the figure), generated by two
$2\,mm$ tilted Calcite crystals with perpendicular optical axes
\cite{Cohen14}. The bit-flip is generated by a $45^o$ half
waveplate (HWP) for the horizontal polarization (sad smiley in the
figure). The error correction is composed of two quarter
waveplates (QWPs) placed in the two paths and a $22.5^o$ HWP in
the noisy path. A PBS completes the error correction protocol. The
multiple phase accumulation part is implemented by a $31.5\,m$ long delay
line, projecting photons from different pulses \cite{EliPRL13}.
Two $22.5^o$ HWPs are placed before the projecting PBS to avoid
the collapse of the state by the measurement of the first photon.
The block of the projecting PBS and the HWPs was not used, as the
results in this work were taken without the multiple phase accumulation
part. Eventually, basis transformation is carried out with HWPs
and QWPs and the photons are detected by single photon detectors.
The data are accumulated by FPGA electronics.}
\end{figure}

The full error correction scheme is presented in Fig.
\ref{ExperimentalSetup}. First, the Bell state
\begin{equation}\label{Original_phip}
 |\phi^+\rangle =\frac{1}{\sqrt{2}}\left( |H\rangle_1|H\rangle_2 + |V\rangle_1|V\rangle_2 \right),
\end{equation}
is generated, where the first qubit, denoted by
$|\cdotp\rangle_1$, is protected and the second, denoted by
$|\cdotp\rangle_2$ and used for the phase measurement, is not.
After the phase measurement the state is
\begin{equation}\label{phi_mes}
 |\phi^{\theta}\rangle =\frac{1}{\sqrt{2}}\left( |H\rangle_1|H\rangle_2 + e^{i\theta}|V\rangle_1|V\rangle_2 \right).
\end{equation}
If the noise is applied, the state becomes a mixture of the
original state and
\begin{equation}\label{phi_err}
 |\psi^{\theta}\rangle =\hat{\sigma}_{I}\otimes\hat{\sigma}_{x}|\phi^{\theta}\rangle =\frac{1}{\sqrt{2}}\left( |H\rangle_1|V\rangle_2 + e^{i\theta}|V\rangle_1|H\rangle_2 \right).
\end{equation}
To correct the error, a phase of $\frac{\pi}{2}$ is added to the
two qubits and a Hadamard transform is operated solely on the
second. Then, the two photons are directed to the two inputs of a
polarizing beam splitter (PBS). Only the events in which the two
output paths of the PBS are both occupied by a photon are
post-selected, thus guaranteeing that both post-selected photons
are either horizontally or vertically polarized. After the error
correction operation, both the perturbed and unperturbed states
return to
\begin{equation}\label{phi_fix}
|\phi^{\theta}\rangle,|\psi^{\theta}\rangle^{\underrightarrow{correction}}|\phi^{\theta}\rangle
=\frac{1}{\sqrt{2}}\left( |H\rangle_{1'}|H\rangle_{2'} +
e^{i\theta}|V\rangle_{1'}|V\rangle_{2'}\right).
\end{equation}
Here the entanglement is between the two spatial paths, $1'$, the
un-delayed spatial path and, $2'$, the delayed spatial path.

Error correction is realized by employing a non-unitary operation
using a PBS and post-selection. This is analogous to applying an
error correction without measurement \cite{Mike_Ike} and
initialization. The difference here is that this is realized
without extending the size of the code, but by applying a decoding
operation.


In order to repeat the measurement, one photon is measured while
its twin (denoted as $|\cdotp\rangle_{2'}^0$) is delayed for time
$\delta$ until another pair of entangled photons,
$\frac{1}{\sqrt{2}}\left(
|H\rangle_1^\delta|H\rangle_2^\delta+|V\rangle_1^\delta|V\rangle_2^\delta
\right)$, is generated, where the upper index denotes the time of
creation. After the second pair accumulates the phase and is
corrected, the delayed photon from the first pair interferes on a
PBS with the un-delayed photon (denoted as
$|\cdotp\rangle_{1'}^\delta$) from the second pair, swapping the
entanglement and resulting in the state $\frac{1}{\sqrt{2}}\left(
|H\rangle_{1'}^0|H\rangle_{2'}^\delta + e^{i2\theta}
|V\rangle_{1'}^0|V\rangle_{2'}^\delta \right)$. After $N-1$
similar iterations the state is
\begin{equation}\label{phi_fix_N}
|\phi^{N\theta}\rangle =\frac{1}{\sqrt{2}}\left(|H\rangle_{1'}^0|H\rangle_{2'}^{(N-1)\delta} + e^{iN \theta} |V\rangle_{1'}^0|V\rangle_{2'}^{(N-1)\delta}\right),
\end{equation}
which oscillates N times faster than the state of single
iteration. Thus, in a scenario in which the gate would have been
realized deterministically, the sensitivity would have been
increased by a factor of $\sqrt{N}$ compared to the shot-noise
limit, the limit of classical measurements. More experimental
details and a discussion on the entanglement swapping between
delayed photons can be found in Ref. \cite{EliPRL13}.

\section{Results}

We divide the demonstration into two parts: correction of bit-flip
errors and the successive accumulation of consecutive measurement
results. The second part has already been demonstrated by our
group in a previous work \cite{EliPRL13}; here we demonstrate the
error correction part.

\subsection{Single phase accumulation}

\begin{figure}[t]
\centering\includegraphics[angle=270,width=86mm]{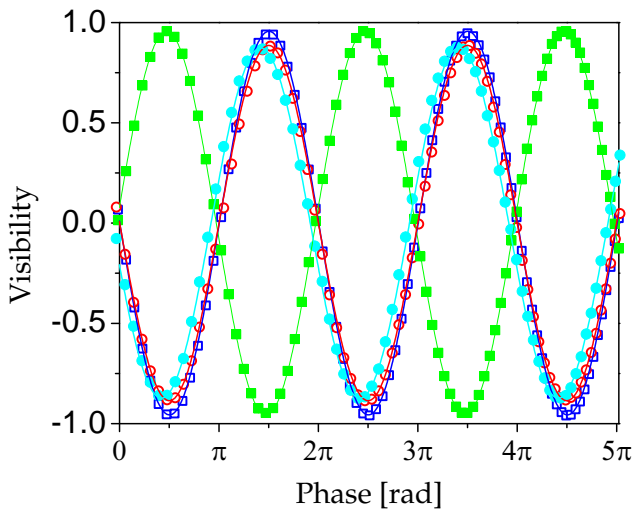}
\caption{\label{visibility}(color online). The visibility $V_{RL}$
as a function of the phase. Solid and empty symbols denote
perturbed and unperturbed states, respectively. Circles and boxes
denote corrected and uncorrected states, respectively. Solid lines
represent fits Sine functions. The corrected states have the same
phase as the original state, while the perturbed uncorrected state
has an opposite phase. Errors were calculated assuming Poisson
error distributions. They are smaller than the symbol sizes, and
thus they are not displayed.}
\end{figure}

The phase information can be retrieved by interfering the $H/V$
amplitudes. One way to achieve this is to detect the photons in
the circular polarization basis. A right (left) handed circularly
polarized photon is defined as $|R\rangle(|L\rangle) =
\frac{1}{\sqrt{2}}(|H\rangle\begin{smallmatrix} +  \\ (-) \end
{smallmatrix} i|V\rangle)$. Figure \ref{visibility} shows the
results of the phase dependent interference, where the visibility
of the state in the circular basis $V_{RL}$ is presented as a
function of the phase. The $V_{RL}$ visibility is defined as
$P_{RR}-P_{RL}-P_{LR}+P_{LL}$, where for example $P_{RL}$ denotes
the probability for coincidence of a right and a left circularly
polarized photon.
A visibility as high as $94\%$ was observed in the case of the
original (unperturbed uncorrected) state, which indicates a well prepared initial state,
where the corrected states have slightly lower visibilities due to
optical imperfections. The phases of the corrected states are
similar to the original phase, but the phase of the perturbed
uncorrected state is not.

\begin{figure}[t]
\centering\includegraphics[angle=270,width=86mm]{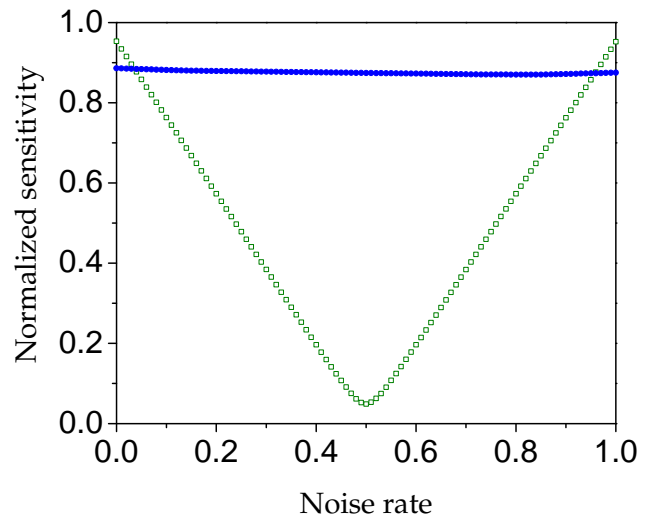}
\caption{\label{sensitivity}(color online). The sensitivity as a
function of the noise rate, with an error correction (solid blue circles)
and without (empty green boxes). Errors were calculated assuming Poisson
error distributions. They are smaller than the symbol sizes, and
thus they are not displayed.}
\end{figure}

Our demonstration provides an important example where using an
entangled (quantum) state has an advantage over a separable
(classical) state \cite{Demkovich14}. The entanglement allows for
the correction of the state and preservation of the measurement of
the phase without an error. In Fig. \ref{sensitivity}, we used the
experimental results presented in Fig. \ref{visibility} to
estimate the sensitivity as a function of the noise rate. The
count number is normalized, making the maximal sensitivity unity.
The sensitivity after the error correction is greater than $0.87$
and nearly independent of the noise rate. The sensitivity without
error correction is linearly impaired and almost vanish as the
noise rate achieves $0.5$. In the presented scenario, in
principle, the correction could restore the fidelity completely as
the sensing of the birefringent phase and the bit-flip noise are
spatially disconnected and are induced in the right order, i.e.,
the signal first and then the bit-flip. Noteworthy is that in
other scenarios in which both the noise and the signal overlap in
time or in space, or occur in different order, a large rate of
error correction is required to reach high sensitivity.
Nevertheless, when the signal measurement and noise stages are
repeated multiple times, the significance of their order vanishes
with the increasing number of stages.

\subsection{Towards multiple phase accumulations}

\begin{figure}[t]
\centering\includegraphics[angle=270,width=86mm]{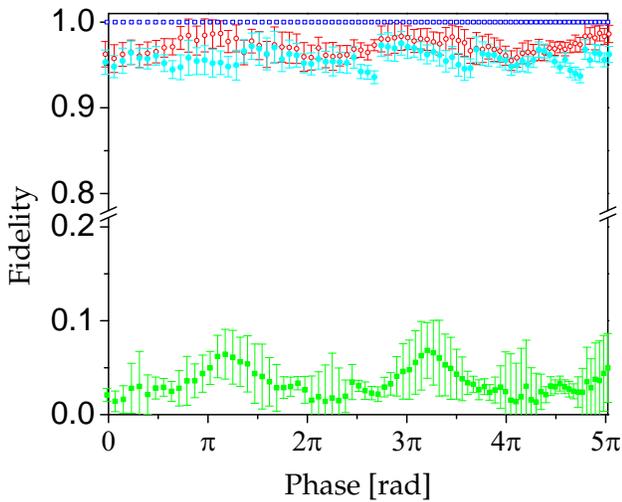}
\caption{\label{fidelity}(color online). The fidelity with the
original state as a function of the phase. The notations and
colors are the same as in Fig. \ref{visibility}. Errors were
calculated assuming Poisson error distributions.}
\end{figure}

As mentioned, the projection on the PBS is used to accumulate the
phase repeatedly. It works only if the projection is between two
$|\phi^{\theta}\rangle$'s, which means that the entire state must
be corrected. Thus, we also performed a complete quantum state
tomography to the perturbed/unperturbed corrected/uncorrected
states for all phases \cite{James01}. We have calculated the
fidelities between all states and the original state \cite{Josza}, which is presented in Fig.
\ref{fidelity}. The fidelities are close to one for all phases
which indicates that we have recovered the original state. In
contrast, the fidelity with the perturbed uncorrected state is
close to zero, demonstrating the orthogonality between these
states.

\section{Conclusions}

In conclusion, we have realized a proposal to use a quantum error
correction protocol to recover sensitivity in the presence of a
single type error such as bit-flipping noise. A linear optics
implementation of the measurement and the error correction is
presented and demonstrated using a PBS and post-selection on the
polarization degree of freedom of photons. The perturbed state was
projected back onto the original state. The results show a
significant sensitivity difference between the corrected and
uncorrected states, where more than $87\%$ of the visibility is
preserved, independently of the noise rate.

\section{Appendix}

\renewcommand{\theequation}{a.\arabic{equation}}

\subsection{General noise}
First we would like to present the error correction protocol for a
more general noise. The noise rotation that occurs with
probability $p$ is
\begin{eqnarray}\label{noised_state}
&&|H\rangle^{\underrightarrow{error}} |\psi\rangle = \alpha |H\rangle +\beta |V\rangle\,,\\
\nonumber &&|V\rangle^{\underrightarrow{error}}
|\psi^{\bot}\rangle = e^{i\epsilon}(\beta^* |H\rangle -\alpha^*
|V\rangle) \,.
\end{eqnarray}
where, without loss of generality, we assume that $\beta>0$ and real.  The
entangled perturbed state is $|\phi^{\theta}\rangle_{noised}
=\frac{1}{\sqrt{2}}\left( |H\rangle_1|\psi\rangle_2 +
e^{i\theta}|V\rangle_1|\psi^{\bot}\rangle_2 \right)\,.$ The
experiment that was realized is a specific example for $\beta=1$
and the phase $\epsilon=0$.

We now prove that the transformation,
\begin{eqnarray}\label{transformation}
\hat{U} = \begin{pmatrix} \cos\eta & e^{i\xi}\sin\eta \\
-e^{-i\xi}\sin\eta & \cos\eta \end{pmatrix}
\end{eqnarray}
on the noisy qubit and the applied post-selection correct the
error, where $ \cos\eta=\frac{\beta}{\sqrt{\beta^2 +
|e^{i\chi}-\alpha|^2}}$ , $\chi = \frac{\epsilon-\pi}{2}$ and
$\xi$ is the phase of the non-zero complex number
$e^{i\chi}-\alpha$. This transformation can be generated in the
lab by a phase shifter, a half-wave plate at angle $\eta$ and
another phase shifter. It can be noticed that
\begin{eqnarray}\label{identities}
&&\langle H | \hat{U} | H\rangle = \langle V | \hat{U} | V\rangle = \cos\eta\,,\\
\nonumber
&&\langle H | \hat{U} | \psi \rangle = \alpha \cos\eta +\beta e^{i\xi}\sin \eta \,,\\
\nonumber
&&\langle V | \hat{U} | \psi^{\bot} \rangle =
-e^{i\epsilon}(\alpha^* \cos\eta +\beta e^{-i\xi}\sin \eta)\,.
\end{eqnarray}
The second and third terms are simplified by taking a common
factor of $\cos \eta$. Then, we replace $\beta \tan \eta$ by
$|e^{i\chi}-\alpha|$ but $\xi$ is the phase of this number. Thus,
$\alpha$ ($\alpha^*$) is cancelled out and we get:
\begin{eqnarray}\label{identities}
    &&\langle H | \hat{U} | H\rangle = \langle V | \hat{U} | V\rangle = \cos\eta\,,\\
    \nonumber &&\langle H | \hat{U} | \rho \rangle = e^{i\chi} \cos \eta \,,\\
    \nonumber &&\langle V | \hat{U} | \rho^{\bot} \rangle = e^{i\chi} \cos \eta \,.
\end{eqnarray}
Where in the third expression the definition for $\chi$ is used.

Next, only the events where the two photons have the same
polarization are post-selected via a PBS. Therefore, $| \psi
\rangle \rightarrow e^{i\chi}\cos\eta | H\rangle$ and $|
\psi^{\bot} \rangle \rightarrow e^{i\chi}\cos\eta | V\rangle$ and
we get
\begin{eqnarray}\label{tran_PS}
&&|\phi^{\theta}\rangle^{\underrightarrow{\hat{U}+postselection}} \cos \eta |\phi^{\theta}\rangle   \,,\\
\nonumber
&&|{\phi^{\theta}\rangle_{perturbed}}^{\underrightarrow{\hat{U}+postselection}}
e^{i\chi} \cos \eta |\phi^{\theta}\rangle \,.
\end{eqnarray}
The state is corrected up to a global phase $\chi$ with a
probability of $\cos^2 \eta$.

\subsection{Full scheme derivation}
Using deterministic gates the setup is composed of a gate that
maps the noise to one qubit and the information to the other,
followed by an initialization of the qubit that contains the noise
\cite{thomas}. A more detailed explanation is needed where a PBS and
post-selection is used, where one can think that the
post-selection for the error correction and entanglement swapping
are mixed. The reason it does work is due to post-selection both
in space and time.

The derivation is divided into three steps: First, we show that
for a four photon event without noise the final state is
\begin{equation}\label{final_stt}
\frac{1}{\sqrt{2}}\left(  |H\rangle_{1'}^0|H\rangle_{2'}^{\delta} + e^{i2\theta} |V\rangle_{1'}^0|V\rangle_{2'}^{\delta} \right)\,,
\end{equation}
where as in the text the upper index represents the time of
creation and the lower index represents the spatial path. Second,
we will generalize it to a four photon event with or without
noise. Third, we will generalize this to a $2N$ photon event.

We assume that two pairs of photons were generated, and we neglect
high order terms. We only look at sequences where one photon is
detected at time zero, two photons are detected at time $t=\delta$
(in different spatial paths) and one photon is detected at time
$t=2\delta$. $\delta$ is the travel time of a photon in the delay
line.
After the measurement and correction of the first pair (i.e. after
the PBS and before the post-selection) the state is:
\begin{eqnarray}\label{phi_before_PS}
&\frac{1}{\sqrt{2}}\big( |H\rangle_{1'}|H\rangle_{2'}+  e^{i\theta}|V\rangle_{1'}|V\rangle_{2'}    \\
\nonumber &+|H\rangle_{2'}|V\rangle_{2'} -
e^{i\theta}|V\rangle_{1'}|H\rangle_{1'} \big)\,.
\end{eqnarray}
The last two terms contribute to two photons at time $t=0$ or
$t=\delta$. If the two photons are measured in different
detectors, the post-selection fails and data are disregarded. If
the two photons are measured with the same detector as one photon,
the sequence will be a three photon event or less, and thus will
be discarded. Therefore, the temporal post-selection only leaves
the two first terms, where one photon will be detected at time
$t=0$ and one at time $t=\delta$. Thus, in order to get the right
sequence, another photon must be detected at time $t=\delta$ and
its twin at time $t=2\delta$. This can only happen if the second
pair is generated at a delay of $\delta$ with respect to the
generation of the first pair and exactly one photon enters the
delay line. If these conditions are fulfilled, the four photon
state is
\begin{eqnarray}\label{Two_phi_after_PS}
&\frac{1}{2}\left(  |H\rangle_{1'}^0|H\rangle_{2'}^\delta + e^{i\theta} |V\rangle_{1'}^0|V\rangle_{2'}^\delta \right)\times\\
\nonumber &\left(  |H\rangle_{1'}^\delta|H\rangle_{2'}^{2\delta} + e^{i\theta} |V\rangle_{1'}^\delta|V\rangle_{2'}^{2\delta} \right)\,.
\end{eqnarray}

The next step is entanglement swapping on a PBS. We first define
$|\phi^{\theta}\rangle = \cos\frac{\theta}{2}|\phi^{+}\rangle
-i\sin\frac{\theta}{2}|\phi^{-}\rangle$ (up to a global phase).
Then, two half-wave plates rotate the basis from $H/V$ to $P/M$,
the $45^\circ$ diagonal linear polarization basis. After returning
to the $H/V$ basis, Eq. \ref{Two_phi_after_PS} becomes
\begin{eqnarray}\label{Two_phi_after_rotation}
\left(\cos\frac{\theta}{2}|\phi^{+}\rangle^{0,\delta}_{1',2'}
-i\sin\frac{\theta}{2}
|\psi^{+}\rangle^{0,\delta}_{1',2'}\right)\times\\
\nonumber
\left(\cos\frac{\theta}{2}|\phi^{+}\rangle^{\delta,2\delta}_{1',2'}
-i\sin\frac{\theta}{2}
|\psi^{+}\rangle^{\delta,2\delta}_{1',2'}\right)\,,
\end{eqnarray}
where the double index refers to the first and second photons in
the Bell state.
We consider only the events where the delayed photon of the first
pair (denoted by $|\cdot\rangle_{2'}^{\delta}$) and the
non-delayed photon of the second pair (denoted by
$|\cdot\rangle_{1'}^{\delta}$) are in different paths after the
PBS, which gives:
\begin{eqnarray}\label{phi_after_PS}
&&|\phi^{+}\rangle^{0,\delta}_{1',2'} |\phi^{+}\rangle^{\delta,2\delta}_{1',2'} \rightarrow|\phi^{+}\rangle^{0,2\delta}_{1',2'}\\
\nonumber &&|\phi^{+}\rangle^{0,\delta}_{1',2'}|\psi^{+}\rangle^{\delta,2\delta}_{1',2'} \rightarrow|\psi^{+}\rangle^{0,2\delta}_{1',2'}\\
\nonumber &&|\psi^{+}\rangle^{0,\delta}_{1',2'}|\phi^{+}\rangle^{\delta,2\delta}_{1',2'} \rightarrow|\psi^{+}\rangle^{0,2\delta}_{1',2'}\\
\nonumber &&|\psi^{+}\rangle^{0,\delta}_{1',2'}|\psi^{+}\rangle^{\delta,2\delta}_{1',2'} \rightarrow|\phi^{+}\rangle^{0,2\delta}_{1',2'}\,.
\end{eqnarray}
Using basic trigonometric identities we get the state, $\cos
{\theta} |\phi^{+}\rangle^{0,2\delta}_{1',2'}-i\sin
{\theta}|\psi^{+}\rangle^{0,2\delta}_{1',2'}$. Another rotation by
a half-wave plate results in the state of Eq. \ref{final_stt}, as
was required.

The generalization to a perturbed state is straightforward. The
error correction post-selection is done by post-selecting the time
of photon detections. This post-selection is independent from the
noise. After this post-selection, the state is corrected back to
$|\phi^{\theta}\rangle $. Thus, the entanglement swapping is
between two corrected states, and the result is the same.

Finally, a similar derivation applies to any number of photon
pairs. An event of one photon in time zero, two photons in time $t
= \delta, 2\delta,..., (N-2)\delta$ and one photon in time
$(N-1)\delta$ can only be detected if all the error correction
post-selections succeeded. Then, the entanglement swapping is
always between two states of $|\phi^{\theta}\rangle $ (with
different $\theta$'s), which results in the state
\begin{equation}\label{last_one}
\frac{1}{\sqrt{2}}\left(
|H\rangle_{1'}^0|H\rangle_{2'}^{(N-1)\delta} + e^{iN\theta}
|V\rangle_{1'}^0|V\rangle_{2'}^{(N-1)\delta} \right)\,.
\end{equation}


\end{document}